\documentclass{article}

\usepackage[pdftex]{graphicx}

\def\eps{\varepsilon}
\def\To{\rightarrow}
\def\ToP{\stackrel{P}{\To}}

\begin{document}

%
%
%
%
%
%
%
%

\setcounter{page}{2}
\vspace*{2\baselineskip}

\begin{center}
  \textbf{\large Modern Sequential Analysis and its Applications to Computerized Adaptive Testing}

\bigskip

\textbf{Jay Bartroff}\\ Department of Mathematics, University of Southern California{\renewcommand{\thefootnote}{}\footnote{Address
 correspondence to  Jay Bartroff, Department of Mathematics, USC, 3620 S Vermont Ave KAP 108, Los Angeles, CA, USA; E-mail: bartroff@usc.edu}}

\medskip

\textbf{Matthew Finkelman}

Department of Biostatistics, Tufts University

\medskip

\textbf{Tze Leung Lai}

Department of Statistics, Stanford University
\end{center}

\bigskip

\noindent\textbf{\small Abstract:} {\small After a brief review of recent advances in sequential analysis
  involving sequential generalized likelihood ratio tests, we discuss
  their use in psychometric testing and extend the asymptotic
  optimality theory of these sequential tests to the case of
  sequentially generated experiments, of particular interest in
  computerized adaptive testing. We then show how these methods can be
  used to design adaptive mastery tests, which are asymptotically
  optimal and are also shown to provide substantial improvements over
  currently used sequential and fixed length tests.}

\vspace{.3in}



\section{1.\ Introduction}
Sequential analysis of data is used in many types of psychometric
tests.  Some of these are computerized adaptive testing, classroom
interaction assessment and intervention, psychological studies
involving longitudinal data, depression diagnosis, and crime-suspect
identification tests.  The purpose of this article is to show how
powerful techniques in modern sequential analysis can be used to
design efficient testing procedures. In particular, we focus on
computerized adaptive testing and show how these techniques can lead
to substantial improvements over previous sequential procedures as
well as conventional tests that do not incorporate early stopping.

Computerized adaptive testing (CAT) has been extensively studied in
the psychometric literature as an efficient alternative to
paper-and-pencil tests.  By selecting an examinee's $k$th test item
based on his/her responses to items $1,\ldots,k-1$, a CAT is tailored
to the individual taking the examination and is thus intended to
quickly home in on each examinee's ability level.  When the test is
designed to measure only one trait, the ability level is typically
denoted by $\theta$ to conform to the notation in standard item
response theory.  There is substantial literature on
efficient estimation of $\theta$ in CAT applications (van der Linden
and Pashley,\nocite{vanderLinden00} 2000; Chang and
Ying,\nocite{Chang03} 2003) and on the problem of classifying
examinees as either masters or non-masters in a given content area
(Reckase, \nocite{Reckase83}1983; Lewis and Sheehan,\nocite{Lewis90}
1990; Chang,\nocite{Chang04} 2004; Chang,\nocite{Chang05} 2005).  The latter problem, known as
computerized mastery testing (CMT), can be formalized by setting a cut
point $\theta_0$ and defining an examinee as a master if and only if
his/her ability level $\theta$ meets or exceeds that cut point.

Typically, a CMT assumes a so-called ``indifference region''
$(\theta_-, \theta_+)$ containing $\theta_0$, which may be thought of
as the ability values which are close enough to the cut point that
neither a decision of mastery nor a decision of non-mastery would
result in a serious error.  The statistical hypothesis of mastery is
then given by $H_0:\theta\ge \theta_+$, while the hypothesis of
non-mastery is given by $H_1:\theta\le\theta_-$.  In a CMT, it is
often the case that an examinee can be quickly identified as a master
or a non-master if that examinee's ability is substantially higher or
lower than the cut point.  Therefore, CMT often involves
variable-length testing whereby the number of items administered
varies by examinee.  An important goal in CMT is to strike a balance
between the confidence of a correct decision and the economy of the
number of items administered.  There are thus two essential components
of any CMT: (i) the stopping rule that determines when to cease
testing and make a classification decision; (ii) the method used to
select items adaptively based on an examinee's item response pattern.

The sequential probability ratio test (SPRT; Wald, \nocite{Wald47}1947) has
been studied as a candidate stopping rule (Spray and Reckase, \nocite{Spray96}1996;
Eggen, \nocite{Eggen99}1999; Vos, \nocite{Vos00}2000; Chang, \nocite{Chang04}2004) for CMT.  The SPRT has shorter average test lengths
than fixed-length tests with the same type I and II error rates at two
specific points along the $\theta$ scale.  Although it has shorter
average length, the SPRT does not constrain the maximum number of
items administered. For a test to have no more than $N$ items, it
is necessary to use a truncated SPRT (TSPRT), which halts testing
and makes a classification decision once $N$ items have been
administered. Suppose that
$k$ items have been presented to an examinee, yielding the responses
$u_1,\ldots,u_k$, where \begin{equation}\label{32} u_i = \left\{
  \begin{array}{ll}
1,& \mbox{if the examinee answers the $i$th item presented correctly}\\
0,& \mbox{if the examinee answers the $i$th item presented incorrectly.}
  \end{array}\right.\end{equation} The classical theory of the SPRT assumes independence
of responses so that the likelihood of $\theta$ is
\begin{equation}\label{37}L_k(\theta)=\prod_{i=1}^k [p_i(\theta)]^{u_i}
[1-p_i(\theta)]^{1-u_i},\end{equation}
where $p_i(\theta)=P_\theta\{u_i=1\}$ for an
examinee of ability $\theta$.  The SPRT stops after the $k$th item and rejects $H_0: \theta\ge\theta_+$
if
\begin{equation}\label{17}\log\frac{L_k(\theta_-)}{L_k(\theta_+)} \ge A,\end{equation}
or accepts $H_0$ if \begin{equation}\label{18}\log\frac{L_k(\theta_-)}{L_k(\theta_+)}\le
-B,\end{equation} where $A,B>0$ are chosen so that $P_{\theta_+}\{\mbox{reject
  $H_0$}\}=\alpha$ and $P_{\theta_-}\{\mbox{accept
  $H_0$}\}=\beta$. Wald's \nocite{Wald47}(1947) approximation yields 
\begin{equation}\label{12}
  A=\log((1-\alpha)/\beta),\quad
  B=\log((1-\beta)/\alpha).\end{equation} The TSPRT stops with
  (\ref{17}) or (\ref{18}) for $k<N$, and if stopping does not
  occur with the $(N-1)$st item, it rejects $H_0$ if and only if
\begin{equation}\label{19}\log\frac{L_N(\theta_-)}{L_N(\theta_+)}\ge
  C.\end{equation} For the TSPRT, Spray and Reckase \nocite{Spray96}(1996) and Eggen
  \nocite{Eggen99}(1999) still use (\ref{12}) for the values of $A$ and $B$ and use for
  (\ref{19}) the value 
\begin{equation}
  \label{14}
  C=(A-B)/2.
\end{equation} The motivation for (\ref{19}) and (\ref{14}) is that all examinees classified as
non-masters at the $N$th item have a log-likelihood ratio no further from $A$ than
$-B$, and those classified as masters have a
log-likelihood ratio no further from $-B$ than $A$. Since
(\ref{12}) is based on the error rates of the
untruncated SPRT, the true error rates of the truncated procedure,
whose decision at truncation is given by (\ref{19}) and (\ref{14}),
are often substantially inflated (see Table~\ref{table2}
below). This is of particular concern in CMT, where $\alpha$
represents the percentage of proficient examinees who are failed.

We address herein this problem by using a new class of stopping rules,
recently introduced in the sequential analysis literature for testing
the composite hypotheses $H_0$ versus $H_1$ subject to type I and II
error probability constraints and a prescribed maximum number of
observations. These tests use the generalized likelihood ratio (GLR)
statistics instead of simple likelihood ratios and have been shown to
have certain optimality properties when the observations are
independent and identically distributed (i.i.d.) and whose common
distribution belongs to an exponential family. In a CAT, the
successive responses $u_1,u_2,\ldots$ of an examinee, however, are not
identically distributed and may not even be independent if the items
are chosen adaptively, since most CATs choose the next item to be an
unused item in the available item pool according to some
criterion. This is also another reason besides the truncation issue
why the theory of the SPRT is not applicable to CMTs. We
show in Section~2.2 that modern sequential testing theory can
in fact accommodate this adaptive feature in sequential experimentation
in addition to providing  efficient stopping and terminal decision
rules. In fact, the methodology developed in Section~2, which is
illustrated by applications to CMTs, is applicable to a large variety
of psychometric tests, allowing sequential choice of experiments
(items in the CMT context) and providing a powerful test at the
conclusion of the study that satisfies the prescribed type I error
probability constraint and whose expected sample size is nearly
optimal and can be considerably smaller than the prescribed maximum
sample size.

This paper is organized as follows. Section~2 first gives a review of
recent developments in sequential GLR tests of composite hypotheses
based on i.i.d.\ observations from an exponential
family. Then the i.i.d.\ assumption is removed and the
theory is extended
to the case where experiments are chosen adaptively to generate an
observation (response) at the next stage. The methodology is then
applied to the design of efficient CMTs, in which the sequential choice of
experiments corresponds to sequential selection of items to be
administered to an examinee based on item response theory. Section~3
reports simulation studies of the performance of the proposed CMT and
compares it with commonly-used fixed-length tests and TSPRTs.
Section~4 gives some concluding remarks.

\section{2.\ Modern Sequential Methods and Their Applications to CMT}

\subsection{2.1 Efficient Sequential GLR Tests for I.I.D. Observations}

To summarize recent advances in sequential hypotheses testing in a
general framework that is applicable to psychometric testing including
CMTs, let $X_1, X_2,\ldots$ be i.i.d.\ observations from
an exponential family of densities $f_\theta(x)=e^{\theta
  x-\psi(\theta)}$ and let $L_k(\theta)$ denote the likelihood
$$L_k(\theta)=\prod_{j=1}^k f_\theta(X_j).$$
The SPRT, which uses
the simple likelihood ratio $\log (L_k(\theta_-)/L_k(\theta_+))$ to test the
hypotheses $H_0:\theta\ge\theta_+$ versus $H_1:\theta\le\theta_-$, is only optimal in the rare case
that $\theta$ is exactly $\theta_-$ or $\theta_+$ and has to allow the
possibility of many more than $N$ observations being taken. A powerful technique in modern sequential analysis that allows the
type I error probability to be controlled while having a maximum
sample size $N$ and preserving asymptotic optimality over the entire
parameter space (instead of just at $\theta_+$ or $\theta_-$) is the modified
Haybittle-Peto test (Lai and Shih, \nocite{Lai04}2004). Let $\widehat{\theta}_k$
denote the maximum likelihood estimator (MLE) of $\theta$ based on
$X_1,\ldots,X_k$. The modified Haybittle-Peto test involves replacing
the simple likelihood ratios in (\ref{17}), (\ref{18}), and (\ref{19})
by the GLR statistic $L_k(\widehat{\theta}_k)/L_k(\theta')$, which ``self-tunes'' to information about
the true $\theta$ accumulating in $\widehat{\theta}_k$ over the course
of the test and in which $\theta'$ denotes the appropriate alternative
that will be specified below. Lai and
Zhang \nocite{Lai94}(1994) and Lai \nocite{Lai97,Lai01} (1997, 2001) have shown that sequential GLRs are
efficient in many testing problems when the thresholds (e.g., $A, B$ in (\ref{17}), (\ref{18})) are appropriately adjusted, even
when $\theta$ is multidimensional. However, the distribution of the
GLR is generally more complicated than the simple likelihood ratio,
and the classical approximations (\ref{12}) do not apply.  But with
the modern computing power that is readily available to practitioners,
Monte Carlo simulation or recursive numerical methods are viable and
often the preferred methods for computing the thresholds, especially in
light of the inflated error probabilities that result from using
classical approximations with truncated tests; see Jennison and
Turnbull \nocite{Jennison00}(2000, Chapter 19) and Lai and Shih \nocite{Lai04}(2004).

The modified Haybittle-Peto test of the hypotheses $H_0: \theta\ge
\theta_+$ and $H_1:\theta\le\theta_-$ can be described as follows. If
$N$ is the maximum number of observations and $\alpha, \beta$ are the
desired type I and II error probabilities, then there is a value
$\theta_-^{(N)}<\theta_+$ such that the likelihood ratio
test of $\theta=\theta_+$ versus $\theta=\theta_-^{(N)}$ based on $N$
observations has type I and II
error probabilities $\alpha$ and $\beta$; in this sense $\theta_-^{(N)}$ is
referred to as the \emph{implied alternative}. Note that $\theta_-^{(N)}$
is not necessarily equal to $\theta_-$, but it is the appropriate
alternative to consider given the parameters $N,\alpha,\beta$, and
$\theta_+$. In addition, focusing on the implied alternative $\theta_-^{(N)}$
frees us from having to specify the alternative $\theta_-$, which
is often chosen arbitrarily in practice. A detailed example of how to compute $\theta_-^{(N)}$ is given in Section~3.1. Let $0<\rho <1$. For $\rho
N\le k<N$,
the modified Haybittle-Peto test stops after the $k$th item and
rejects $H_0$ if
\begin{equation}
  \label{2}
  \widehat{\theta}_k <\theta_+\quad\mbox{and}\quad
  \log\frac{L_k(\widehat{\theta}_k)}{L_k(\theta_+)}\ge A,
\end{equation} or accepts $H_0$ if 
\begin{equation}
  \label{1}
  \widehat{\theta}_k >\theta_-^{(N)}\quad\mbox{and}\quad
  \log\frac{L_k(\widehat{\theta}_k)}{L_k(\theta_-^{(N)})}\ge B,
\end{equation} for some constants $A$ and $B$. 
 For $k=N$, the test is always terminated, with $H_0$ rejected if and
  only if 
\begin{equation}
  \label{3}
  \widehat{\theta}_N <\theta_+\quad\mbox{and}\quad
  \log\frac{L_N(\widehat{\theta}_N)}{L_N(\theta_+)}\ge C
\end{equation} for some constant $C$. If both (\ref{2}) and (\ref{1})
  hold for some $k$ (which can only happen when $A$ and $B$ are
artificially small), then either decision can be made, for example,
always accepting $H_0$ or deciding based on $\widehat{\theta}_k$. In CMT,
where the false negative rate is critical, a simple approach is to
classify as proficient, i.e., accept $H_0$, when this occurs; we take
this as the definition here.

Next the thresholds $A,B$, and $C$ are chosen so
that the false negative error rate does not exceed $\alpha$ and the
false positive error rate, at the alternative $\theta_-^{(N)}$ implied
by the maximum number $N$ of observations, is close to
$\beta$. Specifically, $A$, $B$, and $C$ will be chosen so that 
 \begin{eqnarray}
 &&P_{\theta_-^{(N)}}\{\mbox{(\ref{1}) occurs for some $k<N$}\} =\eps\beta,\label{6}\\
 && P_{\theta_+}\{\mbox{(\ref{2}) occurs for some $k<N$, (\ref{1}) does not
  occur for any $j\le k$}\} =\eps\alpha\label{4}\\
 &&P_{\theta_+}\{\mbox{(\ref{2}), (\ref{1}) do not occur for any $k<N$,
 (\ref{3}) occurs}\} =(1-\eps)\alpha\label{5}
 \end{eqnarray}
 for some $0<\eps<1$.  In practice any value of $\eps$ giving a test
 with desirable properties can be used, and Lai and Shih (2004) have
 shown that values $1/3\le\eps\le 1/2$ work well in a variety of
 settings. The values of $A,B$, and $C$ that satisfy
 (\ref{6})-(\ref{5}) can be determined by Monte Carlo
 simulation, a detailed example of which is given in Section~3.1, or by
 numerical methods based on the following normal approximation to the log-likelihood
 ratios: When $\theta$ is the
 true parameter, \begin{equation}\label{10}Z_k=\mbox{sign}(\widehat{\theta}_k-\theta)
 \left\{2k\log 
  \frac{L_k(\widehat{\theta}_k)}{L_k(\theta)}\right\}^{1/2}\approx
  N(0,k) \end{equation} for large $k$, with
  independent increments $Z_k-Z_{k-1}$ (with $Z_0=0$). The normal approximation (\ref{10}) suggests
replacing the signed-root statistic $Z_k$ by a sum of independent standard
normal random variables $S_k=Y_1+\cdots+Y_k\sim N(0,k)$ so that, for
example, the condition (\ref{2}) becomes $S_k/\sqrt{k}\le-\sqrt{2A}$. Then, in
place of (\ref{6})-(\ref{5}),
$B$, $C$, and $A$ can be successively found by solving  
\begin{eqnarray}
&&P\{S_k/\sqrt{k}\ge\sqrt{2B}\;\;\mbox{for some $k<N$}\}=\eps\beta\label{21}\\
&&P\{S_k/\sqrt{k}\le-\sqrt{2A}\;\;\mbox{for some $k<N$}\}=\eps\alpha\label{22}\\
&&P\{S_k/\sqrt{k}>-\sqrt{2A}\;\;\mbox{for all $k<N$,}\;\;
S_N/\sqrt{N}\le-\sqrt{2C}\}=(1-\eps)\alpha.\label{23}
\end{eqnarray} The left hand sides of (\ref{21})-(\ref{23}) can be
computed by recursive one-dimensional numerical integration; see
Jennison and Turnbull (2000, Chapter 19) for a more detailed
discussion. 

Closed-form approximations to the probabilities in
(\ref{6})-(\ref{5}) have been developed by Siegmund (1985, Chapter 4)
to compute them approximately without using Monte Carlo or numerical
integration. Letting $\phi$ and $\Phi$ be the standard normal density
and c.d.f. and $m_0$ the smallest integer $\ge \rho N$, the normal
approximation (\ref{10}) used in conjunction with Siegmund's
\nocite{Siegmund85}(1985) boundary crossing probability approximation
yields 
\begin{equation}\frac{1}{2}\left[(\sqrt{2A}-1/\sqrt{2A}) \phi(\sqrt{2A})\log
  (N/m_0)+4\phi(\sqrt{2A})/\sqrt{2A}\right]\label{7}\end{equation} 
  as an
  approximation to (\ref{4}), and  
\begin{equation}\label{8}\Phi(-\sqrt{2C})+\frac{\phi(\sqrt{2A})}{\sqrt{2A}}\left[\log\sqrt{N/m_0}-2+A\log(C/A)\right]\end{equation}
  as an approximation to (\ref{5}). 
  The values of $A$ and $C$ can therefore be determined by first
  setting (\ref{7}) equal to $\eps\alpha$ and solving numerically, and
  then setting (\ref{8}) equal to $(1-\eps)\alpha$ and solving
  numerically. Replacing $A$ by $B$ in (\ref{7}) yields an analogous
  approximation for the probability in (\ref{6}), which can be solved
  numerically to find $B$.

  The modified Haybittle-Peto test with thresholds $A,B,C$ satisfying
  (\ref{6})-(\ref{5}) has type~I error rate $\alpha$ and never takes
  more than $N$ observations. It has asymptotically the
  smallest possible sample size of all tests with the same or smaller
  type I and II error probabilities.  This was proved by Lai and Shih
  (2004, Theorem~2(i)) in the context of group sequential tests, and their proof can also be used to establish the following
  ``fully sequential'' version, of particular interest in CAT.

\medskip\textit{Theorem~1.} Let $0<\rho <1$, and let $X_1,X_2,\ldots$ be i.i.d.\
  observations from an exponential family with parameter $\theta$. Let
  $\mathcal{T}_{\alpha,\beta,N}$ be the class of all tests of
  $H_0:\theta\ge\theta_+$ taking no more than $N$ but no fewer than
  $\rho N$ observations and
  with error probabilities not exceeding $\alpha$ and $\beta$ at
  $\theta=\theta_+$ and $\theta_-^{(N)}$, the alternative for which the
  likelihood ratio test of $\theta=\theta_+$ versus $\theta=\theta_-^{(N)}$
  based on $N$ observations has type I and II error probabilities
  $\alpha$ and $\beta$. If $M$ is the sample size of the modified Haybittle-Peto test,
  then as $\alpha\To 0$ and $\beta\To 0$ such that $\log\alpha\sim\log\beta$,
  \begin{equation}\label{11}E_\theta M\sim\inf_{T\in\mathcal{T}_{\alpha,\beta,N}} E_\theta T\end{equation}
  for all $\theta$.

\subsection{2.2  Extension to Sequentially Generated Experiments}

The primary motivation behind CAT is to reduce the length of the test
by adaptively creating a test better suited to the individual
examinee (see Bickel, Buyske, Chang, and Ying, \nocite{Bickel01}2001).  This is accomplished by choosing an examinee's $(k+1)$st
test item based on his/her previous responses $u_1,\ldots,u_k$.
Hence the responses are no longer i.i.d., violating a basic
assumption in Theorem~1 and also in the optimality theory of the SPRT. Another extension of Theorem~1 that needs to be made for applications to CAT is that the exponential family of density functions in Theorem~1 has to be generalized to the form \begin{equation}\label{24}f_{\theta,j}(x)=e^{x\tau_j(\theta)-\psi(\tau_j(\theta))},\quad
j\in J,\end{equation} where $J$ is a set of experiments initially available. In particular, for CAT, whose likelihood function is given in (\ref{37}), 
\begin{eqnarray}
\tau_j(\theta)&=&\log \frac{p_j(\theta)}{1-p_j(\theta)}\label{38}\\
\psi(\tau_j(\theta))&=&\log(e^{\tau_j(\theta)}+1)=-\log[1-p_j(\theta)].\label{39}
\end{eqnarray} Each item $j$ in (\ref{24}) is a reparameterized exponential family, and when the $\tau_j$ are smooth functions of $\theta$, as in (\ref{38}) (provided the $p_j$ are smooth), then the form of the exponential family (\ref{24}) implies that $\psi$ is smooth also. Then the standard formulas for exponential families give
\begin{equation}
E_{\theta,j}X_i=\psi'(\tau_j(\theta)),\quad \mbox{Var}_{\theta,j}X_i=\psi''(\tau_j(\theta)),\label{40}
\end{equation} and therefore \begin{eqnarray}
I_j(\theta,\lambda)&=&E_{\theta,j}\log[f_{\theta,j}(X_i)/f_{\lambda,j}(X_i)]\nonumber\\
&=&\psi'(\tau_j(\theta))[\tau_j(\theta)-\tau_j(\lambda)]-[\psi(\tau_j(\theta))-\psi(\tau_j(\lambda))].\label{41}
\end{eqnarray} Let $j_i$ denote the $i$th sequentially chosen
experiment, and to avoid trivialities from selection rules that somehow look ``into the future,'' we assume experiments are chosen according to some rule that involves only the
previous observations $X_1,\ldots,X_{i-1}$. The
likelihood function still has the form $$L_k(\theta)=\prod_{i=1}^k
f_{\theta,j_i}(X_i)$$ since the
density function of $X_i$ given $X_1,\ldots,X_{i-1}$ is
$f_{\theta,j_i}$. Hence the computation of the error
probabilities, and therefore also of the thresholds $A,B,C$, for the
modified Haybittle-Peto test in the present case can proceed in the
same way as in Section~2.1, in which $|J|=1$ and $f_{\theta,j_i}(X_i)$
is simply $f_\theta (X_i)$.  

Before extending to the CAT setting where each item can only be used once, we extend Theorem~1 to the following setting in which an item can be used multiple times so that its information content can be learned by repeatedly using it, as in the case of nonlinear design of experiments (Fedorov, \nocite{Fedorov97} 1997) and as is used in sequential medical and psychological diagnosis.
 
\medskip\textit{Theorem 2.} Suppose that experiments are
  sequentially chosen from a set $J$ by a rule $\delta$ such that at stage $i$, the choice of
  $j_i$ depends only on $X_1,\ldots,X_{i-1}$, that
\begin{eqnarray}
&& \nu_j=\lim_{n\To\infty} n^{-1}\sum_{i=1}^n
  P\{j_i=j\}\quad\mbox{exists for every $j$,}\label{25}\\
&& \inf_{|\theta|\le a} \sum_{j\in J}\nu_j
  \psi''(\tau_j(\theta))[\tau_j'(\theta)]^2>0\quad\mbox{for all $a>0$,}\label{26}\end{eqnarray} and that the observations follow (\ref{24}). Then (\ref{11}) still holds 
 for all $\theta$,
 where  $M$ is the sample size of the modified Haybittle-Peto test and
  $\mathcal{T}_{\alpha,\beta,N}=\mathcal{T}_{\alpha,\beta,N}(\delta)$ is the class of tests described
 in Theorem~1 that use $\delta$ to select experiments at every
 stage prior to stopping.
 
  \bigskip The proof of Theorem~2 is given in the Appendix, which also
 gives the asymptotic theory of the MLE and GLR statistics in
 sequentially generated experiments from (\ref{24}) under the
 assumptions (\ref{25}) and (\ref{26}). This theory allows us to use
 the approximation (\ref{10}) to compute the probabilities in
 (\ref{6})-(\ref{5}) and thereby determine the thresholds $A,B,C$ of
 the modified Haybittle-Peto test for the general exponential family
 considered here. The assumption (\ref{25}) is a consistency requirement that the long-run frequency $\nu_j$ with which experiment $j$ is used must exist.  For example, if experiments are completely randomized then $\nu_j=1/|J|$ for all $j\in J$. As will be seen in the proof of Theorem~2 in the Appendix, the assumption (\ref{26}) is a uniform convexity requirement of the information numbers (\ref{41}) which can usually be routinely verified. For example, for any finite $J$ satisfying (\ref{38})-(\ref{39}),  (\ref{26}) will hold provided $p_j'(\theta)>0$ for all $j$ and $\theta$ (i.e., the $p_j(\theta)$ are well defined with respect to $\theta$) since, in this case, 
 \begin{equation}
\psi''(\tau_j(\theta))[\tau_j'(\theta)]^2=\frac{[p_j'(\theta)]^2}{p_j(\theta)[1-p_j(\theta)]}\ge 4 [p_j'(\theta)]^2\ge 4\min_{j\in J, |\theta|\le a}[p_j'(\theta)]^2=\gamma>0
\end{equation} for all $j, |\theta|\le a$. Hence, the left-hand-side of (\ref{26}) is at least $\gamma>0$.

We next modify Theorem~2 by imposing the additional constraint that each experiment can be used at most once, as in CAT. This restriction implies that $J=J(N)$ with $|J|\ge N$ and that we cannot learn about an experiment's efficiency directly by using it repeatedly. On the other hand, an experiment's efficiency can be learned indirectly through the estimate $\widehat{\theta}$ of $\theta$.  In particular, suppose that $J$ can be partitioned into a fixed number $K$ of classes $J_1,\ldots,J_K$ with $J_k=J_k(N)$ and such that the experiments in $J_k$ give rise to observations that follow the same distribution in (\ref{24}). That is, assume that \begin{equation}
J=\cup_{k=1}^K J_k\quad\mbox{and for each $k$, there is $\tau^{(k)}$ such that $\tau_j=\tau^{(k)}$ for all $j\in J_k$}. \label{42}
\end{equation}
In practice in CAT, these classes $J_k$ may represent items with the same or similar item response properties. The asymptotic optimality of the modified Haybittle-Peto test in this setting can be proved  by the same arguments as those used in the proof of Theorem~2, provided the classes $J_k$ satisfy some assumptions analogous to (\ref{25})-(\ref{26}). This is the content of the following theorem, whose proof is given in the Appendix.

\medskip\textit{Theorem 3.}  Suppose that $J$ satisfies (\ref{42}) and $|J|\ge N$, that experiments are
  sequentially chosen by a rule $\delta$ such that $j_i$ depends only on $X_1,\ldots,X_{i-1}$, that
\begin{eqnarray}
&& \nu^{(k)}=\lim_{n\To\infty} n^{-1}\sum_{i=1}^n
  P\{j_i\in J_k\}\quad\mbox{exists for every $k=1,\dots,K$,}\label{43}\\
&& \inf_{|\theta|\le a} \sum_{k=1}^K\nu^{(k)}
  \psi''(\tau^{(k)}(\theta))[\tau^{(k)\prime}(\theta)]^2>0\quad\mbox{for all $a>0$,}\label{44}\end{eqnarray} that the observations follow (\ref{24}), and that experiments cannot be used more than once. Then the results of Theorem~2 hold for the modified Haybittle-Peto test, where $\mathcal{T}_{\alpha,\beta,N}(\delta)$ is as described there.
  
  \subsection{2.3  Application: Efficient Design of CMT}

To apply Theorem~3 to the design of efficient CMTs, we use item
response theory (IRT) to model the probability $p_j(\theta)$ that an
examinee of ability $\theta$ gives the correct answer to item $j$. IRT is traditionally utilized in CMT
to provide methods for adaptive item selection as well as to
estimate and compare the respective abilities of examinees who were
administered distinct sets of items. We assume in the sequel the three-parameter
logistic (3-PL) model (Lord, \nocite{Lord80}1980):
\begin{equation}\label{9}
 p_j(\theta) = c_j + \frac{1-
   c_j}{1+e^{-a_j(\theta-b_j)}},\end{equation} with known
   parameters $(a_j,b_j,c_j)$ for all items $j$ in the available item
   pool. 
   
   Any CMT must have an item selection rule as well as a stopping
   rule. This item selection rule is \emph{adaptive} in the sense that
   the choice of the $k$th question for an examinee depends on
   $u_1,\ldots,u_{k-1}$, where the $u_i$ are defined in (\ref{32}) so
   that $P_\theta\{u_i=1|u_1,\ldots,u_{i-1}\}=p_{j_i}(\theta)$, in
   which $j_i$ denotes the item chosen for the $i$th question. Most item selection rules in the literature maximize some
   index of psychometric information at a specified value of $\theta$
   to select the next item for a given examinee.  One such index is
   the Kullback-Leibler (KL) information, which for the 3-PL model (\ref{9}) is
\begin{equation}\label{30}I_j(\theta,\theta') = p_j(\theta)
   \log\frac{p_j(\theta)} {p_j(\theta')} + [1-p_j(\theta)] \log
   \frac{1-p_j(\theta)}{1-p_j(\theta')}.\end{equation} The KL
   information $I_j(\theta,\theta')$ is a measure of the
   distinguishability of the true ability level $\theta$ from level
   $\theta'$ provided by item $j$. Another such measure used in CMT is
   the Fisher 
   information, which for the 3-PL model is $$I_j(\theta)
  =\frac{a_j^2(1-c_j)}{(c_j+e^{a_j(\theta-b_j)})
    (1+e^{-a_j(\theta-b_j)})^2}.$$ 
Reckase \nocite{Reckase83}(1983), Lewis and Sheenan (1990), Spray and Reckase (1996), and Chang and Ying (2003) use procedures that choose the next item in a test to be the
    unused item that maximizes the Fisher information at the cut point
    $\theta_0$ or at a current estimate of $\theta$, like the MLE
    $\widehat{\theta}_k$. Spray and
  Reckase (1996) suggest
  maximizing information at $\theta_0$ rather than $\widehat{\theta}_k$
  when using the SPRT.  Eggen's (1999)
  simulations showed that KL information outperforms both of these approaches based on Fisher
  information in some settings. These adaptive item selection rules
  satisfy (\ref{43})-(\ref{44}) as discussed following Theorem~2.

\section{3.\ Simulation Studies}

\subsection{3.1  Simulation of Proposed CMT}

In this section we compare the fixed-length, TSPRT, and modified Haybittle-Peto tests
of $H_0:\theta\ge\theta_+$ versus $H_1:\theta\le\theta_-$ about the
ability level $\theta$ in the 3-PL model. To isolate the effects of
the different stopping rules, all tests use the same criterion --
maximum Fisher information -- to sequentially choose items. To simulate
the tests in a realistic setting, we used a real item pool from the Chauncey Group International, a subsidiary of the Educational
Testing Service. The pool has 1136 3-PL item
parameters, with $a_j$ ranging between 0.289 and 2.372 and having a median of 0.862, $b_j$ ranging between -5.531 and 5.426 and having a median of -0.943, and $c_j$ ranging between 0.048 and 0.529 and having a median of 0.232. The real-life cut point associated with the item pool
is $\theta_0=-1.32$.  Mimicking simulations by Lin and Spray \nocite{Lin00}(2000),
$\theta_-$ and $\theta_+$ are taken to be $\theta_0\mp .25=-1.07$,
$-1.57$. Following Spray and Reckase (1996), $N$ was set to
50. Table~\ref{table2} gives the type I error probability and the average length of the TSPRT for various values of $\alpha=\beta$.

\begin{center}INSERT TABLE~\ref{table2} ABOUT HERE\end{center}

As mentioned above, the TSPRT with thresholds (\ref{12}) and
(\ref{14}) usually has type I and II error probabilities substantially larger
than the nominal values $\alpha$ and $\beta$.  Table~\ref{table2} shows that the actual type I error probability is roughly constant at about .16 for
$\alpha\le .1$. This is because the thresholds
$A=B=\log((1-\alpha)/\alpha)$ are large enough that truncation occurs
for nearly every examinee, evident through the large average test
lengths, and consequently a large proportion of the examinees are
misclassified at the truncation point. Since the type I error
probability in CMT is the percentage of proficient $\theta_+$-level
examinees who are misclassified as non-proficient, we propose a
modification of the TSPRT by choosing $C$ suitably to make
the type I error probability approximately equal to the nominal value
$\alpha$, rather than use (\ref{14}).  Table~\ref{table3} contains the
average test length and percentage of examinees classified as
non-master, i.e., the power, of the following tests at various values
of $\theta$: The TSPRT using thresholds (\ref{12}), (\ref{14}) with
$\alpha=\beta=.05$; the TSPRT, modified in the way described above
(denoted by modTSPRT), with the same values of 
$A=B=\log(.95/.05)$ but with $C=1.4$ to give type I error
$\alpha=.05$; the fixed-length test with $N=50$, using
classification rule (\ref{19}) with $C=1.28$ that is chosen to give type I
error probability $\alpha=.05$; the modified
Haybittle-Peto test (denoted by modHP) with $A=3.7, B=3.3, C=1.4$ that
are chosen to satisfy (\ref{6})-(\ref{5}) with
$\alpha=\beta=.05$, $\eps=1/2$, $\rho N=5$, and $\theta_-^{(N)}=-1.95$
where the fixed-length test has power $1-\beta=.95$; details of how $A, B, C$ and $\theta_-^{(N)}$ were computed are given below. All four tests choose the next item to be the unused one that
maximizes Fisher information at the MLE, when it exists.  When the MLE does not exist, the tests use Fisher information at the real-life cut point $\theta=\theta_0$. The average test length and
power are computed at eleven values of $\theta$ between $-2$ and $-.5$,
including $\theta_0$, $\theta_+$~(in bold), $\theta_-$, and $\theta_-^{(N)}$ from
10,000 simulated tests each.

\begin{center}INSERT TABLE~\ref{table3} ABOUT HERE\end{center}

The fixed, modTSPRT, and modHP tests have very similar power functions
for $\theta\le\theta_+$. The TSPRT has high power but also greatly
inflated type I error probability $16.1\%$, resulting from the use of
the approximations (\ref{17}), (\ref{18}) in its definition, as
discussed above. The modTSPRT has the same average test length as the
TSPRT because they use the same thresholds $A$ and $B$, and both
provide savings in test test length over the fixed-length test,
particularly at ability levels outside the indifference region
$(\theta_-,\theta_+)$.  The modHP test provides substantial savings in
test length over the fixed length as well as the TSPRT and modTSPRT.
The self-tuning nature of the GLR allows modHP to dramatically shorten
the tests of proficient examinees ($\theta\ge\theta_+$), for whom
modHP is about half the length of modTSPRT. Moreover, the modHP tests
are shorter on average even when $\theta=\theta_+$ or $\theta_-$,
suggesting that the method of computing thresholds (\ref{6})-(\ref{5})
contributes to its efficiency as well as the use of the GLR statistic.

The parameters $A, B, C$ and $\theta_-^{(N)}$ of modHP were computed by Monte Carlo simulation as follows.  Two simple numerical routines were used: Routine~1, which takes as input a candidate value $\theta_-^{(N)}<\theta_+$ and returns the estimated type II error probability of the fixed-length level-$\alpha$ likelihood ratio test of $\theta=\theta_+$ vs.\ $\theta=\theta_-^{(N)}$ of length $N$, based on 10,000 simulated tests using the Chauncey item pool; and Routine 2, which takes as inputs boundaries $A,B$ and $C$ in (\ref{2})-(\ref{3}), $\theta_-^{(N)}$, and the true ability level $\theta$, and returns the average test length and estimated type I error probability of modHP using Fisher information with maximum length $N=50$, based on 10,000 simulated tests using the Chauncey item pool. Routine~1 works by first solving for the critical value $C$ in (\ref{19}) (with $\theta_-$ replaced by $\theta_-^{(N)}$) giving type I error probability $\alpha=.05$. This is done by finding two values of $C$ whose corresponding type I error probabilities bracket .05 and then using the bisection method. After $C$ has been found, the type II error probability is output by simulating tests at $\theta=\theta_1^{(N)}$ and bracketing and bisection are again used to find the value $\theta_-^{(N)}$ giving type II error probability $\beta=.05$.  The value $\theta_-^{(N)}=-1.95$ was found in this way. Next, Routine~2 was used to find $A,B$ and $C$ in (\ref{2})-(\ref{3}) by first setting $A=C=\infty$ and $\theta=\theta_-^{(N)}$ and finding $B$ that satisfies (\ref{6}) with $\eps\beta=.05/2=.025$ by bracketing and bisection. $B=3.3$ was found in this way and used to next find $A$ satisfying (\ref{4}) with $\eps\alpha=.05/2=.025$ by simulating at $\theta=\theta_+$. $A=3.7$ was found and used to similarly find $C=1.4$ satisfying (\ref{5}) with $(1-\eps)\alpha=.05/2=.025$. Both Routines~1 and 2 are stable and run quickly, and $A, B, C$ and $\theta_-^{(N)}$ are computed in a matter of minutes.

\subsection{3.2  Simulation of Proposed CMT with Exposure Control and
  Content Balancing}

Even though the example in Section~3.1 utilizes a real item pool, the tests are compared under
somewhat ideal circumstances where items can be selected purely due to
their statistical properties.  However, since the modified
Haybittle-Peto test presented above relies on no specific item selection rule or IRT
model, it has the flexibility to incorporate  additional
constraints on item selection that arise in typical CATs, such as
exposure control and content balancing in the choice of items. In this
section we illustrate this by presenting a second simulation
study comparing the modified Haybittle-Peto test with the TSPRT and
fixed length test, all using the following simple method for
exposure control and content balancing.

Suppose that the exposure of the items in the pool needs to be controlled
so that each item is administered to no more than a proportion $\pi$
of examinees, on average.  Suppose also that the content of the test
needs to be balanced in the sense that each item in the pool falls
into one of $s$ categories, and these categories should be represented approximately
in given proportions $q_1,\ldots,q_s$, where $\sum_1^s q_i=1$. A
simple way of satisfying these constraints when using a test of
maximum length $N$ is the following. From each category $i=1,\ldots,s$,
first select the $Nq_i/\pi$ (neglecting rounding) items with the largest Fisher
information at the cut-point $\theta_0$, then randomly select $Nq_i$ items from among these, resulting in a new item
pool of
$\sum_1^s Nq_i=N$ items, the proportion $q_i$ of which are in category $i$.  The chance that a given item in category $i$
appears in the new pool is clearly no greater than $Nq_i/(Nq_i/\pi)=\pi$. If a test that allows early stopping is being used, like the
TSPRT or modified Haybittle-Peto test, then the method of
spiraling (Kingsbury and Zara, \nocite{Kingsbury89}1989) can be used so that the category
proportions are close to $q_1,\ldots,q_s$ even when early stopping
occurs; spiraling simply entails choosing at the $(k+1)$st stage an
item from the category $i$ whose proportion in the first $k$ items
differs the most from $q_i$. 

\begin{center}INSERT TABLE~\ref{table4} ABOUT HERE\end{center}

Table~\ref{table4} contains the average test length and power of the
fixed-length ($N=50$), TSPRT, modified TSPRT, and modified Haybittle-Peto tests using
this method of exposure control and content balancing, for various values
of $\theta$ (with $\theta_+$ in bold). For this study, the Chauncey item pool used in
Section~3.1 was randomly divided into $s=3$ ``content'' categories,
$\pi$ was set to .25, and $q_1=.4$, $q_2=.3$, $q_3=.3$. Each entry in
Table~\ref{table4} was computed from 10,000 simulated tests. The fixed-length ($N=50$) test uses
classification rule (\ref{19}) with $C=1.33$, chosen to achieve type I
error probability about $\alpha=.05$. The TSPRT uses the stopping rule
(\ref{17})-(\ref{14}) with $\alpha=\beta=.05$, and the modified TSPRT
(denoted by modTSPRT) uses the same values of $A$ and $B$ but with
$C=1.3$ to ensure type I error probability of $\alpha=.05$, as
discussed in Section~3.1. The modified
Haybittle-Peto test (denoted by modHP) uses $A=3.7$, $B=3.8$,
$C=1.47$ that are chosen to satisfy (\ref{6})-(\ref{5}) with
$\alpha=\beta=.05$, $\eps=1/2$, $\rho N=5$, and
$\theta_-^{(N)}=-2.11$, where the fixed-length test has power
$1-\beta=.95$; these parameters were computed using Monte Carlo simulation similar to the last paragraph of Section~3.1. The tests show very similar relative performance to
those in Table~\ref{table3}.  The modTSPRT and modHP tests
have power functions very similar to the fixed-length
test, while the TSPRT is over-powered, including an inflated type I error probability
of 19.3\% that results
from use of the approximations (\ref{17}), (\ref{18}), (\ref{14}) in
its definition, as discussed above. The modHP tests are substantially
shorter than the TSPRTs for all values of $\theta$ considered, and
particularly for $\theta\ge\theta_+$ where the reduction was around
40\% to 50\%. Note that the tests in
Table~\ref{table4} are less powerful and on average longer than the
corresponding ones in Table~\ref{table3}; this is because they do not always
choose the most informative item available in order to satisfy the exposure control and content balancing constraints. 

\section{4.\ Conclusion}
This paper shows how efficient sequential tests that use
``self-tuning'' sequential GLR statistics can be extended from the
i.i.d.\ setting to incorporate sequentially designed experiments. The
tests are also sufficiently general to handle practical issues that
arise in computerized adaptive testing applications, like the method
used in Section~3.2 to satisfy the constraints on exposure control and
content balancing or the more complex methods proposed by Sympson and
Hetter \nocite{Sympson85}(1985) and Stocking and Swanson
\nocite{Stocking93}(1993). These tests have potential applications in
psychometric testing with sequentially generated experimental designs
and data-dependent stopping rules, as illustrated in Sections~2.3 and
3 for CMT.


\appendix
\renewcommand{\theequation}{A\arabic{equation}}
\setcounter{equation}{0}
\renewcommand{\thesection}{\Alph{subsection}}
\setcounter{section}{0}
\section*{Appendix: Proof of Theorems~2 and 3 and Related Asymptotic Theory}


In order to prove Theorem~2, we modify the basic arguments of Lai and
Shih (2004) that prove
Theorem~1, and whose key ingredients are the following.

(a) Hoeffding's \nocite{Hoeffding60}(1960) lower bound for the expected sample size
$E_\theta T$ of a test that has error probabilities $\alpha$ and
$\beta$ at $\theta=\theta_+$ and $\theta_-^{(N)}$, which simplifies
asymptotically to 
\begin{equation}\label{20}E_\theta (T)\ge (1+o(1))|\log\alpha
  |/\max\{I(\theta,\theta_+),I(\theta,\theta_-^{(N)})\}\end{equation} as
   $\log\alpha\sim\log\beta$, where $I(\theta,\lambda)=E_\theta
   \{\log[f_\theta (X_i)/f_\lambda
   (X_i)]\}=(\theta-\lambda)\psi'(\theta)-\{\psi(\theta)-\psi(\lambda)\}$
   is the Kullback-Leibler information.

(b) The sample size $N$ of the fixed-sample-size likelihood ratio test
of $\theta=\theta_+$ versus $\theta=\theta_-^{(N)}$ with error probabilities
$\alpha$ and $\beta$ at $\theta=\theta_+$ and $\theta_-^{(N)}$, which satisfies
\begin{equation}\label{27}N\sim |\log\alpha |/I(\theta^*,\theta_+)\end{equation} as
$\log\alpha\sim\log\beta$, where $\theta_-^{(N)}<\theta^*<\theta_+$ is the
unique solution of
$I(\theta^*,\theta_+)=I(\theta^*,\theta_-^{(N)})$. Moreover,
$\max\{I(\theta,\theta_+), I(\theta,\theta_-^{(N)})\}$ attains its minimum
at $\theta=\theta^*$.

(c) $\lim_{n\To\infty} P_\theta\{\max_{\rho n\le m\le n}
|\widehat{\theta}_m-\theta|\ge a \} =0$ for every $a >0$.

\bigskip To extend this to Theorem~2, we need analogs of (a), (b), and
(c) to hold for the case of sequentially generated
experiments. Without assuming the $X_i$ to be independent, Lai \nocite{Lai81}(1981, Theorem~2) has derived a
Hoeffding-type lower bound which in our case takes the form 
\begin{equation}\label{28}E_\theta (T)\ge (1+o(1))|\log\alpha
  |/\max\left\{\sum_{j\in J}\nu_jI_j(\theta,\theta_+), \sum_{j\in
   J}\nu_jI_j(\theta,\theta_-^{(N)})\right\},\end{equation} where
   $\nu_j=\lim_{n\To\infty} n^{-1}\sum_{i=1}^nP\{j_i=j\}$ exists by
   (\ref{25}) and $I_j(\theta,\lambda)$ is given by (\ref{41}). Lai's \nocite{Lai81}(1981, Theorem~2) bounds are derived for
   sequential tests of $H_0:P=P_0$ versus $H_1:P=P_1$ with type I and
   type II error probabilities $\alpha$ and $\beta$, based on random
   variables $X_1,X_2,\ldots$ from a distribution $P$ such that $(X_1,\ldots,X_m)$ has joint
   density function $p_m(x_1,\ldots,x_m)$, under the assumptions that for $k=0,1$, 
\begin{eqnarray}
\label{31}&&
   n^{-1}\log[p_n(X_1,\ldots,X_n)/p_{n,k}(X_1,\ldots,X_n)]\quad \mbox{converges
   in probability to $\eta_k$,}\\
\label{29}
   && \lim_{n\To\infty} P\left\{\max_{m\le
   n}\log[p_m(X_1,\ldots,X_m)/p_{m,k}(X_1,\ldots,X_m)]\ge
   (1+\delta)n\eta_k\right\} =0\;\;\mbox{for every $\delta>0$},\end{eqnarray} where
   $p_{n,k}$ denotes the joint density function under $H_k$, $k=0,1$. These conditions hold in the present case, for which \begin{equation}\label{45}
\log L_n(\theta)=\sum_{i=1}^n X_i \tau_{j_i}(\theta)-\sum_{i=1}^n \psi(\tau_{j_i}(\theta))
\end{equation} by (\ref{24}), and $\eta_0=\sum_{j\in J}\nu_j I_j(\theta,\theta_+)$ and $\eta_1^{(N)}=\sum_{j\in J}\nu_j I_j(\theta,\theta_-^{(N)})$, which can be shown by the following argument. We shall use $P$ to denote the probability measure under which the true parameter value is $\theta$, and $\ToP$ to denote convergence in probability under this measure. The notation $o_P(1)$ wll be used to denote a random variable $Y_n$ such that $Y_n \ToP 0$. From (\ref{25}) it follows that $$n^{-1}\sum_{i=1}^n I_{j_i}(\theta,\theta_+)\ToP \sum_{j\in J} \nu_j I_j(\theta,\theta_+)=\eta_0,$$ and combining this with (\ref{40}) and the law of large numbers applied to (\ref{45}) yields 
\begin{eqnarray*}
n^{-1}\log(L_n(\theta)/L_n(\theta_+))&=&n^{-1}\sum_{i=1}^n\{\psi'(\tau_{j_i}(\theta))[\tau_{j_i}(\theta)-\tau_{j_i}(\theta_+)]-[\psi(\tau_{j_i}(\theta))-\psi(\tau_{j_i}(\theta_+))]\}+o_P(1)\\
&=&n^{-1}\sum_{i=1}^n I_{j_i}(\theta,\theta_+)+o_P(1)\ToP \eta_0,
\end{eqnarray*} where we have used (\ref{41}) for the second equality. The same argument can be used to show that $$n^{-1}\log(L_n(\theta)/L_n(\theta_-^{(N)}))-\eta_1^{(N)}\ToP 0.$$ By Taylor's expansion of (\ref{41}), 
   \begin{eqnarray}
I_j(\theta,\lambda)&=& (\theta-\lambda)^2\psi''(\tau_j(\lambda))[\tau_j'(\lambda)]^2/2+o(\theta-\lambda)^2\quad\mbox{for fixed $\lambda$,}\\
I_j(\theta,\lambda) &=& (\lambda-\theta)^2\psi''(\tau_j(\theta))[\tau_j'(\theta)]^2/2+o(\lambda-\theta)^2\quad\mbox{for fixed $\theta$.} \end{eqnarray} Hence the assumption (\ref{26}) guarantees uniform convexity of the information numbers, which can be used in conjunction
with (\ref{25}) to show that (c) still holds in the setting of
Theorem~2. Moreover, modification of the proof
of Theorem~3 and equation~(15) in Lai and Shih (2004) can be used to
show that as $\log\alpha\sim\log\beta$, $N\sim
|\log\alpha|/\sum_{j\in J} \nu_j I_j(\theta^*,\theta_+)$ analogous to (\ref{27}), where
$\sum_{j\in J} \nu_j I_j(\theta^*,\theta_+) =\sum_{j\in J} \nu_j
I_j(\theta^*,\theta_-^{(N)})$, and that
\begin{equation}\label{36}E_\theta M\sim|\log\alpha |/\max\left\{\sum_{j\in J}\nu_jI_j(\theta,\theta_+), \sum_{j\in
   J}\nu_jI_j(\theta,\theta_-^{(N)})\right\}\sim
   \inf_{T\in\mathcal{T}_{\alpha,\beta,N}(\delta)}E_\theta T,\end{equation} proving
   Theorem~2.
   
   Theorem~3 is proved analogously by replacing the ``items'' in Theorem~2 by the ``item classes'' $J_k$ satisfying (\ref{42}).  The conditions (\ref{43})-(\ref{44}) provide the analogs to (\ref{25})-(\ref{26}).
   
   In the sequel we let $\theta_0$ denote the true parameter value to
   study the asymptotic properties of the MLE $\widehat{\theta}_m$ and
   the GLR statistics in sequentially designed experiments that
   satisfy (\ref{25}) and (\ref{26}). Note that (c) ensures that with probability approaching 1,
   $\widehat{\theta}_m$ is near $\theta_0$ for all $\rho n\le m\le n$.
   A standard argument involving martingale central limit theorems
   (Durrett, \nocite{Durrett05}2005, p.\ 411) and Taylor's expansion
   of $\log L_m(\theta)$ around $\theta_0$ can
   be used to show that as $n\To\infty$
\begin{equation}\left\{n\sum_{j\in J}\nu_j
   I_j(\widehat{\theta}_n,\theta_0)\right\}^{1/2}(\widehat{\theta}_n-\theta_0)\quad\mbox{has
   a limiting standard normal distribution,}\label{35}\end{equation}
   and that the signed-root likelihood ratio statistics in (\ref{10}),
   with $\theta$ replaced by $\theta_0$, are asymptotically normal
   with independent increments, generalizing (\ref{10}) from the
   i.i.d.\ case to sequentially generated experiments.


\def\cprime{$'$}

\newpage

\begin{table}[htdp]
\caption{Type I error probability $P_{\theta_+}\{\mbox{reject $H_0$}\}$
  and average test length $E_{\theta_+}T$ of TSPRT using Chauncey item
  pool with maximum test length $N=50$ and thresholds $A=B=\log((1-\alpha)/\alpha))$, $C=0$.}
\begin{center}
\begin{tabular}{c|rccccc}
&$\alpha=$ .001&.005&.010&.050&.100&.200\\
\hline 
$P_{\theta_+}\{\mbox{reject $H_0$}\}$&.165&.163&.163&.161&.165&.193\\
$E_{\theta_+}T$&50.0&50.0&49.7&44.2&36.7&22.4
\end{tabular}
\end{center}
\label{table2}
\end{table}%

\begin{table}[htdp]
\caption{Average test length and power (in parentheses) of the
  fixed-length, TSPRT, modified TSPRT (modTSPRT), and modified Haybittle-Peto (modHP) tests using the Chauncey item pool.}
\begin{center}
\begin{tabular}{r|cccc}
$\theta$&fixed&TSPRT&modTSPRT&modHP\\
\hline
$-0.50$               &50.0 (0.04\%)&27.6 (0.23\%)&27.6 (0.04\%)&12.6 (0.13\%)\\
$-0.75$               &50.0 (0.28\%)&34.7 (1.53\%)&34.7 (0.24\%)&15.8 (0.46\%)\\
$-1.00$               &50.0 (2.70\%)&42.6 (10.5\%)&42.6 (2.57\%)&22.1 (3.27\%)\\
$\theta_+\;\;= -1.07$     &\textbf{50.0 (5.00\%)}&\textbf{44.2 (16.1\%)}&\textbf{44.2 (5.00\%)}&\textbf{24.5 (5.00\%)}\\
$-1.25$               &50.0 (17.5\%)&46.5 (39.0\%)&46.5 (17.1\%)&30.5 (17.1\%)\\
$\theta_0\;\;\;= -1.32$&50.0 (25.5\%)&46.6 (49.2\%)&46.6 (24.0\%)&32.4 (25.1\%)\\
$-1.50$               &50.0 (51.7\%)&44.2 (75.6\%)&44.2 (49.7\%)&35.0 (49.1\%)\\
$\theta_-\;\;=-1.57$      &50.0 (62.9\%)&42.3 (83.3\%)&42.3 (60.3\%)&34.7 (59.4\%)\\
$-1.75$               &50.0 (83.3\%)&36.3 (94.8\%)&36.3 (82.6\%)&30.5 (80.4\%)\\
$\theta_-^{(N)}=-1.95$&50.0 (95.0\%)&29.3 (99.0\%)&29.3 (93.5\%)&23.6 (92.2\%)\\
$-2.00$               &50.0 (95.4\%)&27.8 (99.3\%)&27.8 (94.3\%)&22.1 (93.2\%)
\end{tabular}
\end{center}
\label{table3}
\end{table}

\begin{table}[htdp]
\caption{Average test length and power (in parentheses) of the
  fixed-length, TSPRT, modified TSPRT (modTSPRT), and modified Haybittle-Peto
  (modHP) tests with exposure control and content balancing.}
\begin{center}
\begin{tabular}{r|cccc}
$\theta$&fixed&TSPRT&modTSPRT&modHP\\
\hline
$-0.50$               &50.0 (0.03\%)&36.7 (0.45\%)&36.7 (0.00\%)&17.6 (0.14\%)\\
$-0.75$               &50.0 (0.43\%)&42.1 (2.71\%)&42.1 (0.31\%)&21.7 (0.64\%)\\
$-1.00$               &50.0 (3.04\%)&46.7 (14.1\%)&46.7 (3.27\%)&28.0 (3.12\%)\\
$\theta_+\;\;= -1.07$     &\textbf{50.0 (5.00\%)}&\textbf{47.5 (19.2\%)}&\textbf{47.5 (5.00\%)}&\textbf{29.8 (5.00\%)}\\
$-1.25$               &50.0 (14.7\%)&48.4 (39.8\%)&48.4 (15.4\%)&34.3 (13.9\%)\\
$\theta_0\;\;\;= -1.32$   &50.0 (21.1\%)&48.3 (49.1\%)&48.3 (21.6\%)&35.7 (20.3\%)\\
$-1.50$               &50.0 (40.9\%)&47.0 (71.8\%)&47.0 (42.9\%)&38.1 (38.0\%)\\
$\theta_-\;\;=-1.57$      &50.0 (50.1\%)&46.1 (78.9\%)&46.1 (51.4\%)&37.9 (47.5\%)\\
$-1.75$               &50.0 (72.0\%)&42.7 (91.1\%)&42.7 (73.0\%)&36.4 (67.3\%)\\
$-2.00$               &50.0 (91.2\%)&35.9 (98.3\%)&35.9 (91.7\%)&30.2 (86.2\%)\\
$\theta_-^{(N)}=-2.11$&50.0 (95.0\%)&33.1 (99.2\%)&33.1 (95.3\%)&27.3 (91.2\%)
\end{tabular}
\end{center}
\label{table4}
\end{table}


%




\end{document}